\documentstyle[preprint,aps]{revtex}
\begin{document}
\bibliographystyle{plain}
\title{
Integrability and coherence of hopping
between 1D correlated electrons systems\footnote{Submitted to Phys. Rev.
Letters}
}
\vskip0.5truecm
\author {Fr\'ed\'eric Mila and Didier Poilblanc}
\vskip0.5truecm
\address{
      Laboratoire de Physique Quantique, Universit\'e Paul Sabatier\\
      31062 Toulouse (France)\\
      }
\maketitle

\begin{abstract}
We present numerical evidence that the hopping of electrons between chains
described by the $t-J$ model is coherent in the integrable cases ($J=0$ and
$J=2$) and essentially incoherent otherwise. This effect is {\it not} related
to the value of the exponent $\alpha$, (which is restricted to
the interval [0,1/8] when $0\le J\le 2$), and we propose that enhanced
coherence is characteristic of
integrable systems.
\end{abstract}
\vskip .1truein

\newpage

After the proposal by Anderson that two-dimensional models of correlated
electrons might behave as Luttinger liquids\cite{anderson}, i.e. as their
one-dimensional
analogs, a lot of work has been devoted to studying the effect of a small
transverse hopping $t_\perp$ between two chains. The first conclusion that
has been
obtained by
several authors\cite{note2} is that a small transverse hopping is a relevant
perturbation as long as the exponent $\alpha$
describing the power-law singularity of the momentum distribution in the
isolated chains is smaller than 1, suggesting that a 2D system of coupled
Hubbard
chains flows toward a Fermi liquid fixed point, $\alpha$ being at most 1/8 in
that case. More recently\cite{chakra1}, another point of view has been
emphasized according to
which what really matters is not whether $t_\perp$ is relevant, but whether it
induces a coherent hopping between the chains. A study of the short time
dynamics of two coupled Luttinger liquids by Clarke, Strong and
Anderson\cite{clarke}
suggests that
coherence can be destroyed even if $t_\perp$ is relevant. Within
their approximation, the hopping between chains becomes incoherent for
$\alpha=1/2$, whereas $t_\perp$ remains relevant until $\alpha=1$.
While this result opens new perspectives in the problem of coupled chains, this
analysis suggests that the key
parameter is still $\alpha$, which somehow reflects the strength of the
correlations within the chains. In a very recent paper about a related problem
in other systems, Chakravarty and Rudnick\cite{chakra2} suggest that
incoherence is actually a
generic feature, and that coherence is limited to a very narrow range of
parameters. The origin of this coherence is left as an open issue.

In this Letter, we identify integrability as a very important factor of
coherence between chains. More specifically, we show that the short time
dynamics of the hopping of electrons between two chains is coherent if the
model describing each chain is integrable, that this coherence is destroyed
if one goes away from integrable points, and that this effect cannot be
understood in terms of the exponent $\alpha$.
The system we have studied consists of two chains (so called $2\times L$
ladder)
described by the $t-J$ model
and coupled by a perpendicular hopping $t_\perp$. The Hamiltonian can be
written
\begin{eqnarray}
H=\sum_{\alpha=1,2} \left( -t \sum_{i,\sigma}
P_G^\alpha(c^{\alpha\dagger}_{i\sigma}
c^{\alpha}_{i+1\sigma}+h.c.)P_G^\alpha
+ J \sum_i \vec{S}^\alpha _i .\vec{S}^\alpha _{i+1} \right)
- t_\perp \sum_i (c^{1\dagger}_{i\sigma} c^{2}_{i\sigma} + h.c. )
\end{eqnarray}
where $P_G^\alpha$ is the Gutzwiller projection operator that excludes double
occupancy on chain $\alpha$.
For small J/t ratios (say $J/t < 0.5$) an additional small transverse
exchange coupling $J_\perp=J (\frac{t_\perp}{t})^2$ has also been included
in order to improve the equivalence with
the large-U Hubbard ladder in this parameter regime.
In the isotropic regime ($t_\perp=t$) transverse coherence
can be established \cite{ladder} but hereafter
we restrict ourselves to the small $t_\perp$ regime.
Following Clarke et al, our analysis of coherence
is based on the probability $P(\tau)$ that a system comes back after some
time $\tau$ to its initial state
if one initially puts more particles on one chain than on the other. More
precisely,
$P(\tau)$
is defined by
\begin{eqnarray}
P(\tau) & = & |A(\tau)|^2\\
A(\tau) & = & \big<\psi_0|e^{i(H-E_0)\tau}|\psi_0\big>
\end{eqnarray}
where $|\psi_0\big>$ is the lowest energy ($E_0$)
eigenstate of the system with $t_\perp$ set to
zero and with $\Delta N_e$ more particles on one chain than on the other, while
$H$ is the full Hamiltonian. There
are actually two such states because the excess particles could be on either
chain. In the following, we have taken the symmetric combination of these
states
for numerical convenience\cite{note1}.

In the case of non-interacting particles (referred to hereafter as $U=0$)
$P(\tau)$ exhibits an oscillatory behavior characteristic of coherent
transverse
motion. More precisely,
$P(\tau)=\left|\cos^{\Delta N_e}(t_\perp\tau) +
i^{\Delta N_e} \sin^{\Delta N_e}(t_\perp\tau)\right|^2$
showing a characteristic period $\pi/(4t_\perp)$ for all the excess particles
to
exactly move in phase from one chain to the next\cite{note1}.
In contrast, when interaction between particles is switched on we expect that
$P(\tau)$
never reaches 1 for finite non-zero $\tau$ although an oscillatory behavior
can still occur in the case of coherent transverse hopping.

Since this issue cannot be adressed by perturbative methods
exact diagonalizations of small ladder clusters have been performed.
Using Lanczos technique, it is easier to calculate first the Fourier
transform of $A(T)$ defined by
\begin{eqnarray}
A(E)=-{1 \over \pi} {\rm Im} \big<\psi_0|{1 \over
E-H+E_0+i\epsilon}|\psi_0\big>
\end{eqnarray}
which can be obtained through a continued fraction expansion. This function is
itself quite useful because coherence shows up as symmetric peaks around $E=0$,
while an incoherent system is expected to have a broad peak centered around 0.
We have calculated these quantities for a system of 2 chains with $L=16$ sites
each,
and for a total number of particles $N_e=8$.
For clarity, most of the results
presented below correspond to $\Delta N_e=8$, i.e. to an initial state having
all
the particles on the same chain, but
the main conclusions of this work do not depend dramatically on the value of
$\Delta N_e$ as discussed later.
Note that antiperiodic boundary conditions have been used in the chains
direction
to ensure closed shell fillings of the corresponding non-interacting systems.
Numerical data obtained for $t_\perp/t=0.2$, $\Delta N_e=8$ and for various
values of $J/t$ ranging from 0
to 2.4 are presented in Figures 1, 2 and 3.

The first important result is that the behaviour depends dramatically on the
value of $J/t$. Two typical cases have emerged:

i) $A(E)$ is  a smooth
distribution centered
around $E=0$, and $P(\tau)$ decreases monotonically with t. This is for
instance
what we have obtained for $J/t=0.25$ (see Figs. 1a and 2b). In this case, the
probability that the system goes back to its initial state is continuously
decreasing, which corresponds to a totally incoherent dynamics.

ii) $A(E)$ has three
narrow peaks, one at $E=0$, the other two symmetric around $E=0$, and $P(\tau)$
exhibits damped oscillations. This situation is best exemplified by
$J/t=2$ (see Figs. 1b and 3). In that case, the probability for the system to
go
back to its initial state reaches again substantial values after going down to
almost 0, at least for moderate values of $T$, and the short time dynamics is
coherent. Very regular oscillations have also been observed for $J/t=0$, with a
smaller amplitude than for $J/t=2$ though (see Fig. 2b).

The basic properties of a single chain described by the $t-J$ model are known
from the work of
Ogata et al\cite{ogata}. Since we have 8 particles on $N=2\times 16$ sites, the
relevant band-filling is n=1/8 (4 particles on each chain). Then, going for
$J/t=0$ to $J/t=2$, the exponent $\alpha$ decreases monotonically from 1/8 to
0\cite{note}. Beyond that point, a gap opens in the spin sector, although the
exact location of the critical value is not known. Finally, a phase separation
occurs at $J/t\simeq 2.7$. Comparing this behavior with  our results (coherence
for $J/t=0$ and $2$, incoherence for  $J/t=0.25$), we conclude that the
coherence we have detected cannot be related in any simple way to the exponent
$\alpha$.

We finish the discussion of the data by a comment on the actual role of the
parameter $\Delta N_e$.
So far we have considered the case $\Delta N_e /N_e = 1$ where the initial
state at $\tau =0$
corresponds somehow to a macroscopic perturbation of the system (with full
Hamiltonian H)
far from its equilibrium state.
Fig. 4 shows a comparative study of $P(\tau)$ at $J/t=0.25$ for $\Delta N_e=8$
and $\Delta N_e=2$.
Although $\Delta N_e=2$ corresponds to a much smaller deviation from the
absolute ground state
(with an excitation energy
of only $\sim 0.7 t$ compared to $\sim 4.3 t$ for $\Delta N_e=8$) it
nevertheless gives rise to a
qualitatively similar behavior for $P(\tau)$ showing again no coherence in the
transverse motion.

A natural question which arises then is what causes the special behavior
observed only for $J/t=0$ and $J/t=2$.
The answer we propose is that these are the only two points for which the $t-J$
model is integrable. For $J/t=0$, the model is equivalent to the finite energy
sector of the infinite $U$ Hubbard model, which is known to be soluble by Bethe
ansatz since the work of Lieb and Wu \cite{lieb}, while for $J/t=2$ an
additional supersymmetry
again makes the model integrable by Bethe ansatz \cite{bares}. The remarkable
differences
in Fig. 3 between the results for $J/t=2$ and those obtained for
$J/t=1.6$ and $J/t=2.4$ support this point of view. For both non integrable
cases,
the oscillations are much smaller in amplitude and involve several frequencies,
a sign of a much less coherent dynamics. Coherence might be
completely lost for arbitrary deviation from the supersymmetric case provided
that
the system size is large enough.

So the main conclusion of this work is that the short-time dynamics of
hopping of electrons between chains
is remarkably coherent if the
model describing the chains is integrable and more or less incoherent
otherwise.

Let us now compare our results with those obtained previously by other authors.
The main conclusion of Clarke et al\cite{clarke}, namely that incoherence
can be achieved for
$\alpha < 1$, is confirmed by our results for $J/t=0.25$, which corresponds to
$\alpha \simeq 0.1$. Note that the necessary condition $t_\perp \gg
2\pi\frac{\Delta N_e}{L}(v_c-v_s)$ for the observation of coherence proposed by
Clarke et al is
clearly satisfied for $J/t=0.25$ because the charge and spin velocities $v_c$
and $v_s$ are
nearly equal for that particular value of $J/t$. However, as already been
pointed out,
our results show that coherence does not seem to be linked with the
value of $\alpha$.
More importantly it seems also that the analysis of
the $\tau^2$
term in the expansion of $P(\tau)$ for short times is not sufficient to study
coherence. In Fig. 2a, we have depicted the short time behavior of $P(\tau)$
for 3
extreme cases: $U=0$ (totally coherent), $J/t=0$ (partially coherent) and
$J/t=0.25$ (totally incoherent). The curvatures at $\tau=0$ are
indistinguishable.

The main conclusion of Chakravarty and Rudnick\cite{chakra2}, namely that
incoherence is the
generic behavior, and coherence is somehow accidental, is also in agreement
with
our results. To go beyond this general statement, one should understand what
integrability means in terms of the models they have studied, and this is not
clear yet.

Finally the reason why integrability and coherence are related remains an open
issue. Integrability is already known to have dramatic
consequences on the level statistics\cite{montambaux}, the distribution being
Poisson instead of
Wigner in that case, and on the finite temperature conductivity\cite{castela}.
While the first effect can be qualitatively understood in terms of
level repulsion, a good explanation of the second one is also missing.
More work
is clearly needed to understand the influence of integrability on the dynamical
properties of correlated electrons.

D.P. acknowledges support from the EEC Capital and Mobility program
under grant CHRX-CT93-0332. We also thank IDRIS (Orsay) for
allocation of CPU time on the C94 and C98 CRAY supercomputers.

\begin{figure}
\caption{
Spectral function $A(E)$ for a $2\times 16$ t-J ladder with $N_e=\Delta N_e=8$.
Energies and parameters are measured in unit of t (i.e. $t'=t_\perp/t$).
(a) Non-integrable case $J=0.25$ (and $J_\perp=0.01$); (b) Supersymmetric case
$J=2$.
}
\end{figure}

\begin{figure}
\caption{
Probability $P(\tau)$ vs time $\tau$
calculated at $J/t=0$ and $J/t=0.25$ for $\Delta N_e=8$ and a transverse
coupling $t'=t_\perp/t=0.2$.
The non-interacting case is also shown (thin dashed line) for comparison.
Time is measured in unit of the inverse hopping integral 1/t.
(a) Small time region; (b) Enlarged time interval.
}
\end{figure}

\begin{figure}
\caption{
Probability $P(\tau)$ vs time $\tau$
calculated at the supersymmetric point $J/t=2$ and in its vicinity
for $\Delta N_e=8$. The non-interacting case is also shown and time is measured
in unit of 1/t.
}
\end{figure}

\begin{figure}
\caption{
Probability $P(\tau)$ vs time $\tau$
calculated at $J/t=0.25$ and for both $\Delta N_e=2$ (full line) and $\Delta
N_e=8$ (dashed line).
}
\end{figure}


\begin{references}
\bibitem{note1}  In \cite{clarke} the initial state was not symmetrized
leading to characteristic time scales larger than ours by a factor 2.
\bibitem{note} In fact, it probably vanishes slighlty before $J/t=2$, but this
is irrelevant for the present discussion.
\end{references}
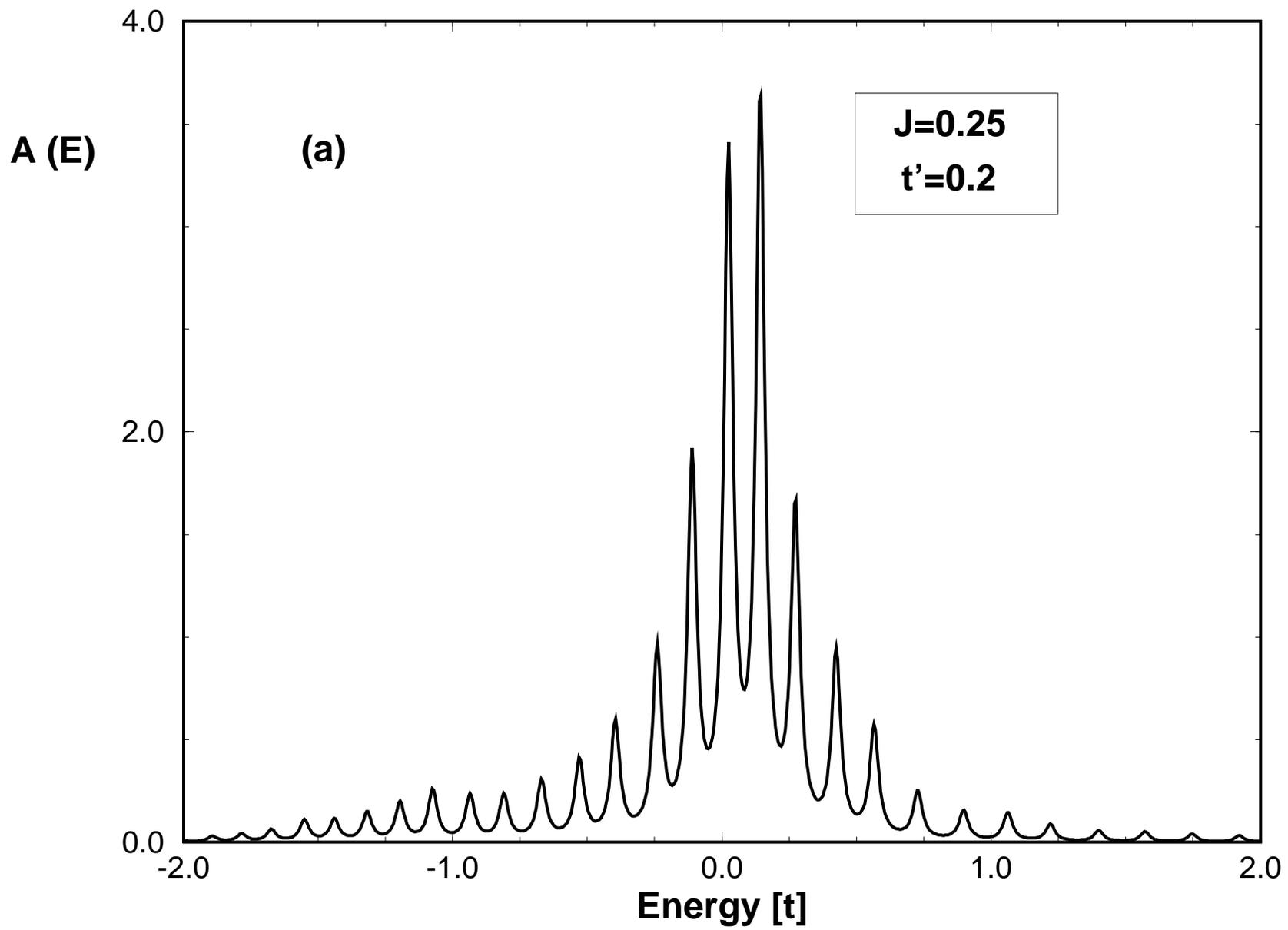
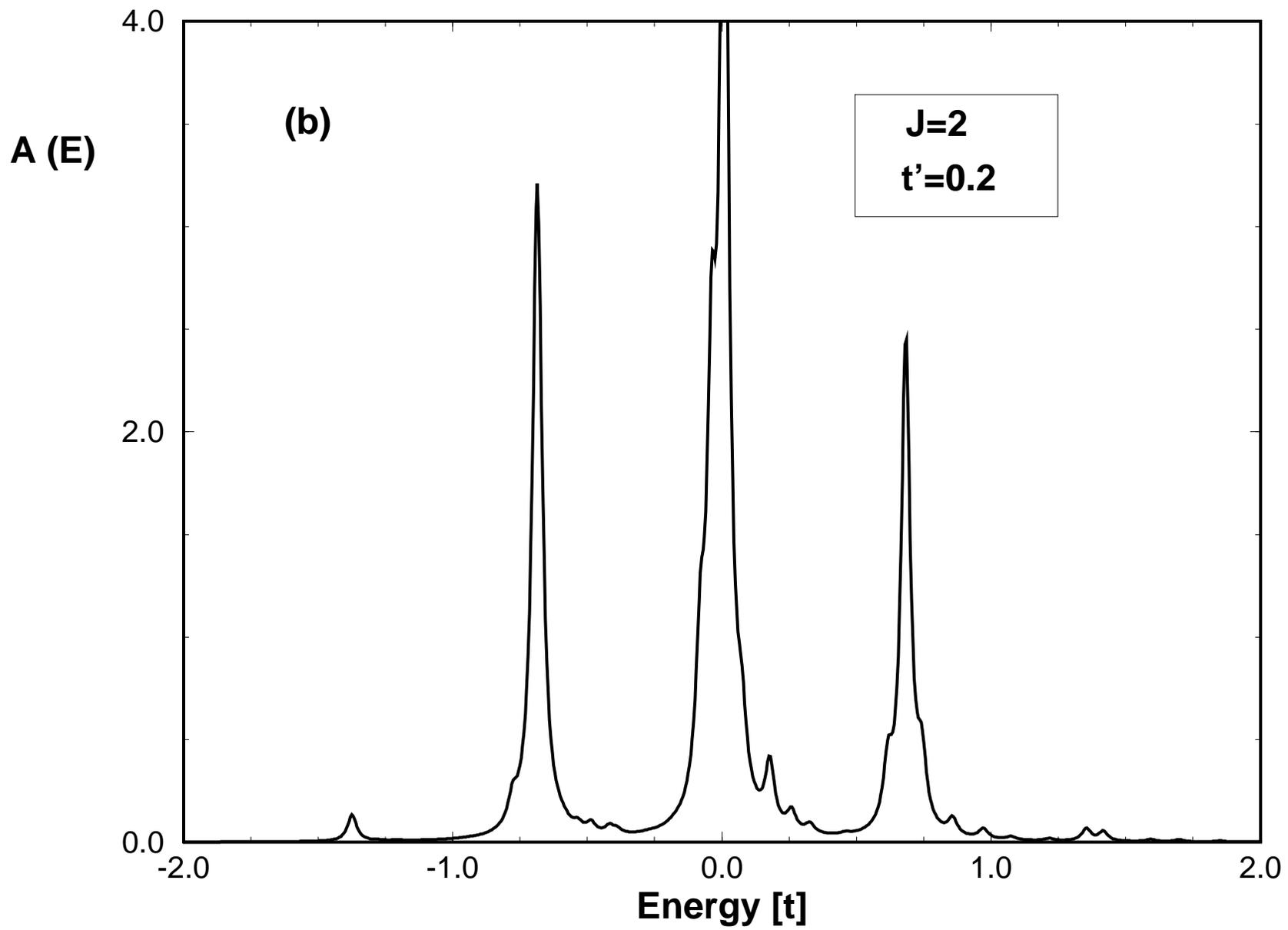
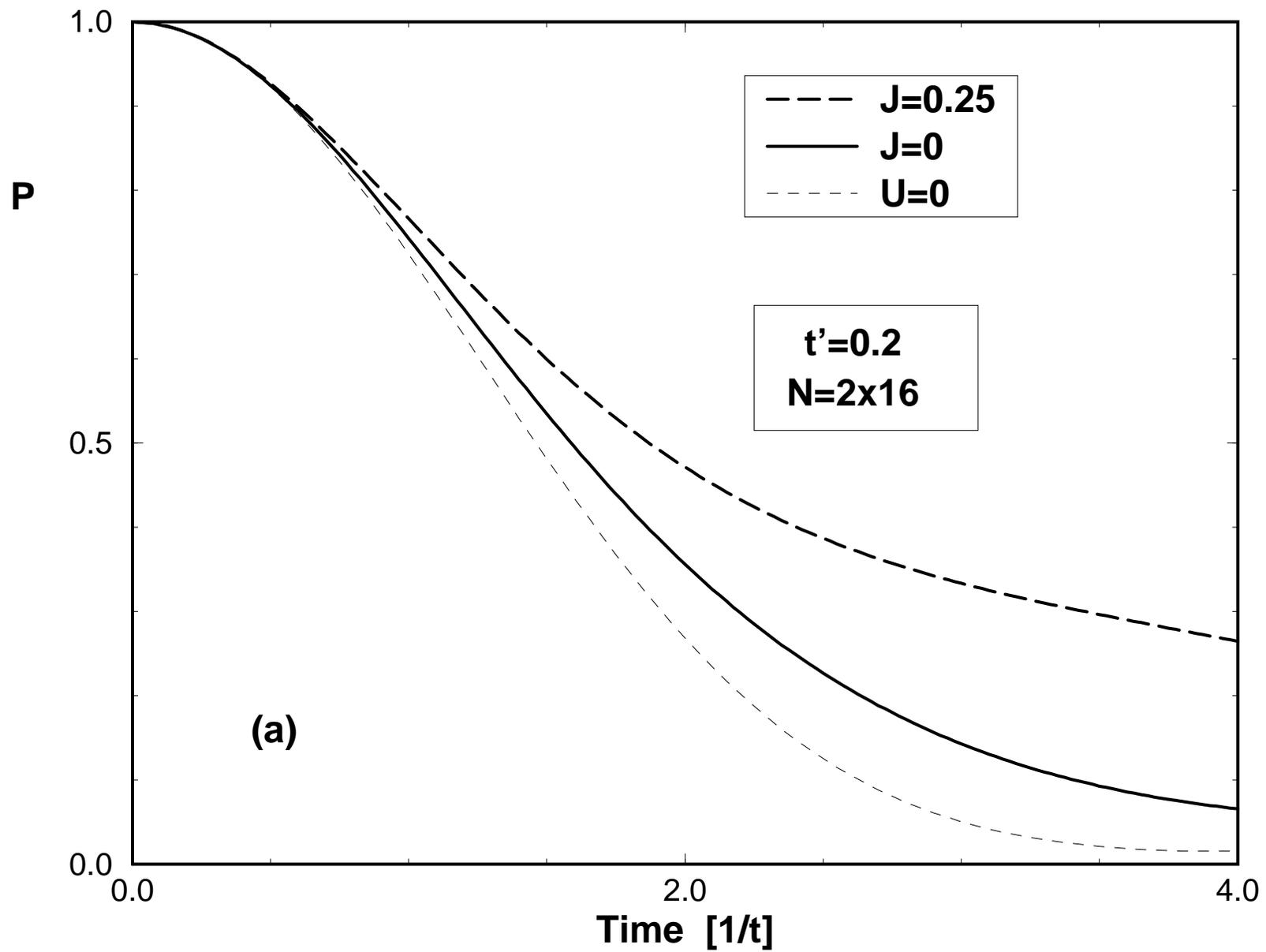
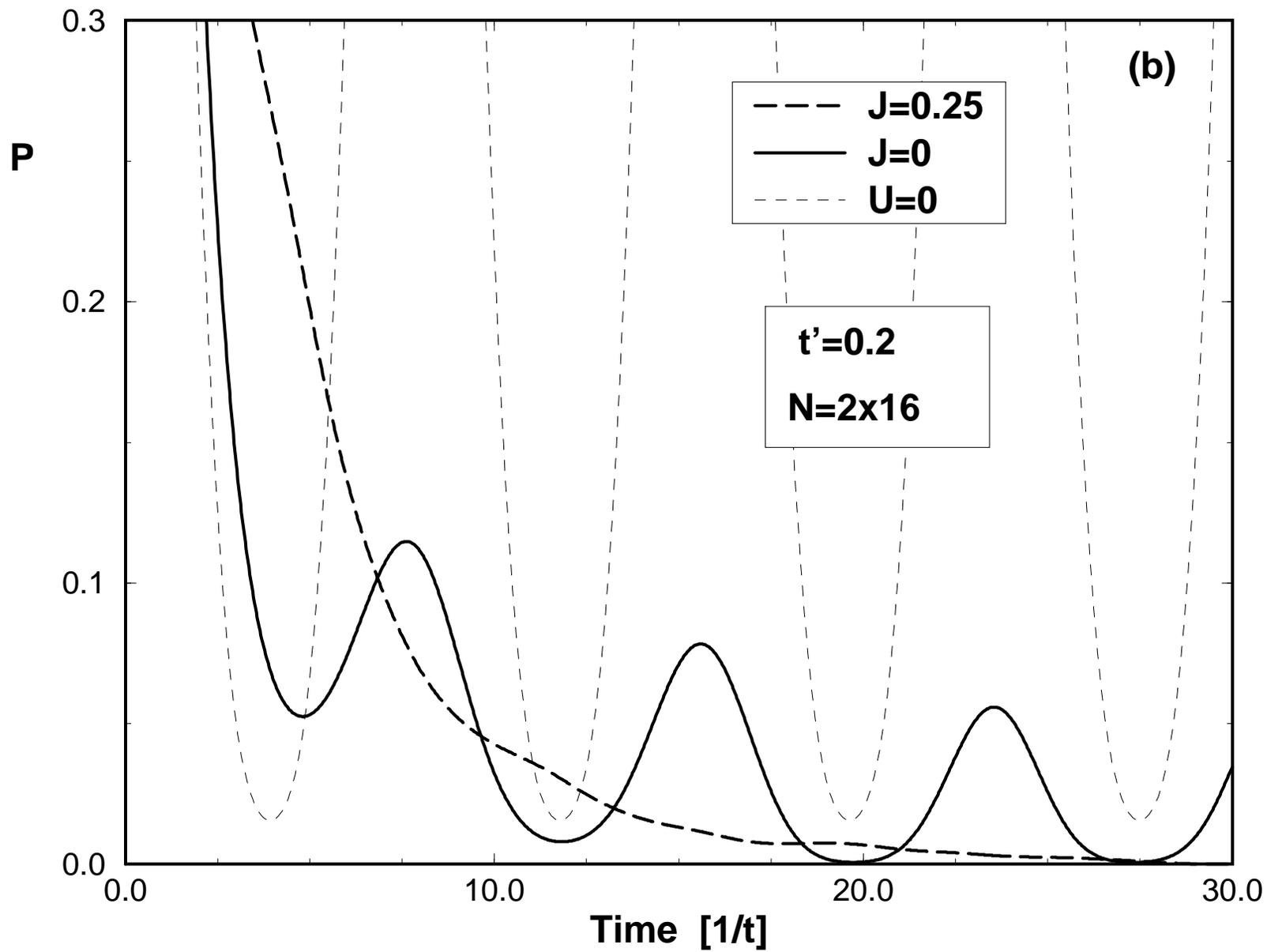
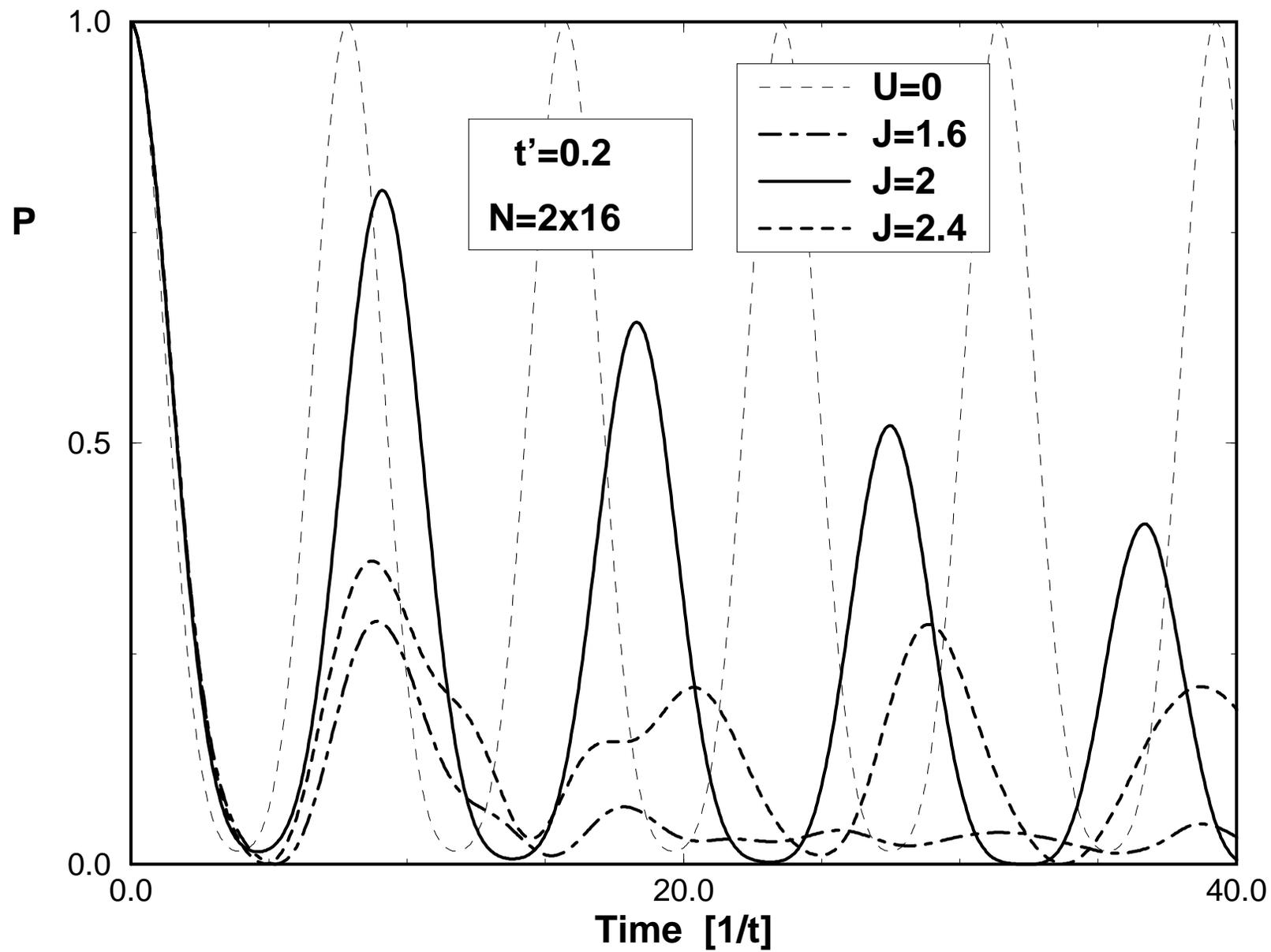
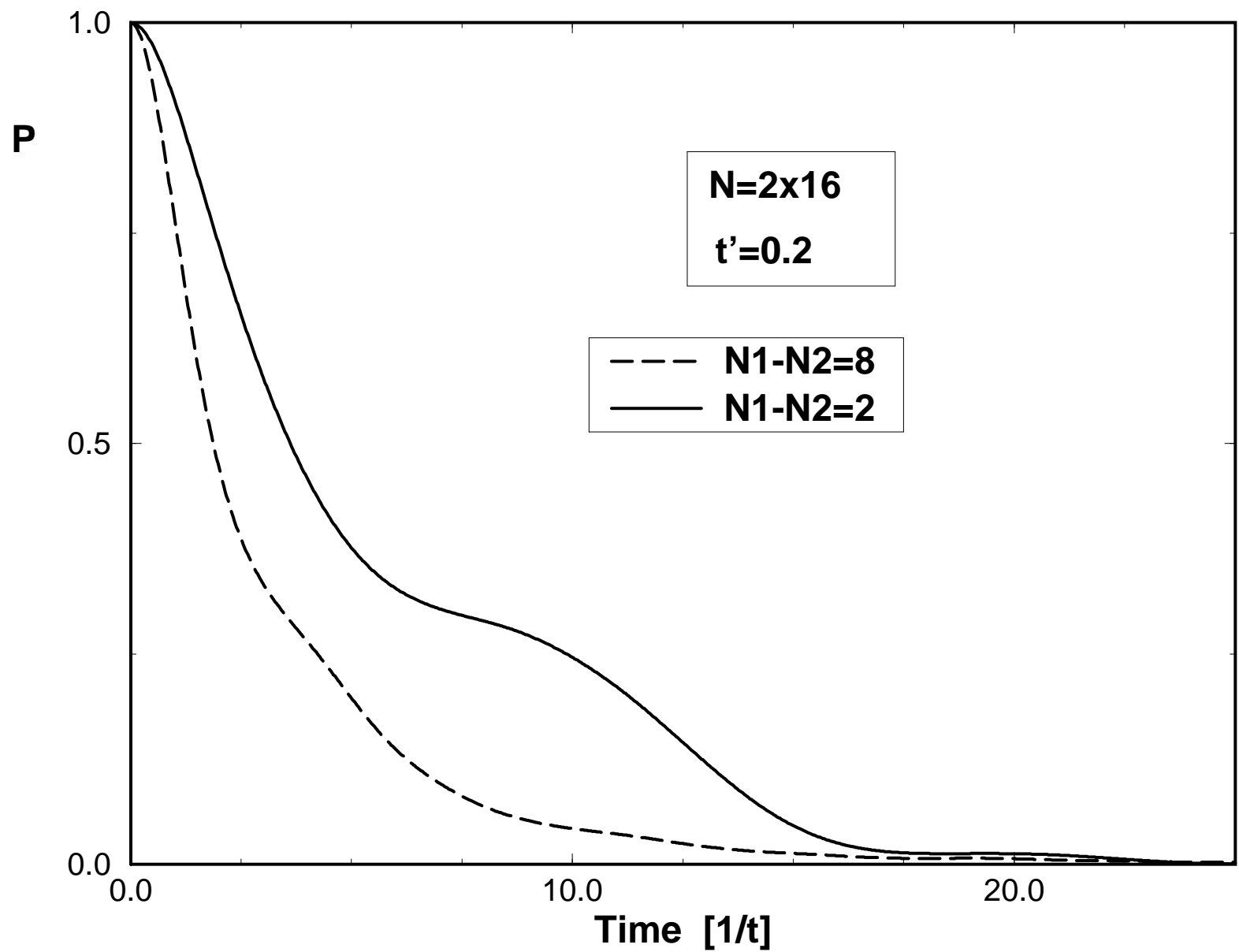
\end{document}